\documentclass[journal]{IEEEtranTICPS}
\usepackage{graphicx}
\usepackage{cite}
\usepackage{picinpar}
\usepackage{amsmath}
\usepackage{url}
\usepackage{flushend}
\usepackage[latin1]{inputenc}
\usepackage{colortbl}
\usepackage{soul}
\usepackage{multirow}
\usepackage{pifont}
\usepackage{color}
\usepackage{alltt}
\usepackage[hidelinks]{hyperref}
\usepackage{enumerate}
\usepackage{siunitx}
\usepackage{breakurl}
\usepackage{epstopdf}
\usepackage{pbox}
\usepackage{subfigure}
\usepackage{algorithm} 
\usepackage[noend]{algpseudocode}
\usepackage{tikz}
\usetikzlibrary{fit,calc}
\newcommand*{\tikzmk}[1]{\tikz[remember picture,overlay,] \node (#1) {};\ignorespaces}
\newcommand{\boxit}[1]{\tikz[remember picture,overlay]{\node[yshift=0pt,fill=#1,opacity=.25,fit={(A)($(B)+(.92\linewidth,\baselineskip)$)}] {};}\ignorespaces}
\newcommand{\boxitt}[1]{\tikz[remember picture,overlay]{\node[yshift=0pt,fill=#1,opacity=.25,fit={(A)($(B)+(.855\linewidth,\baselineskip)$)}] {};}\ignorespaces}
\colorlet{gray}{black!50}
\colorlet{orange}{yellow!60}
\newcolumntype{C}[1]{>{\centering\let\newline\\\arraybackslash\hspace{0pt}}m{#1}}
\usepackage{amssymb}
\usepackage{bm}
\usepackage{booktabs} % use toprule, midrule in Table

\title{Scalable Multi-Agent Reinforcement Learning for Residential Load Scheduling under Data Governance}

\begin{document}


\author{Zhaoming Qin,~\IEEEmembership{Graduate Student Member, IEEE}, Nanqing Dong, Di Liu, Zhefan Wang, and Junwei Cao,~\IEEEmembership{Senior Member, IEEE}
% First A. Author, \IEEEmembership{Fellow, IEEE}, Second B. Author, and Third C. Author, Jr., \IEEEmembership{Member, IEEE}
% \thanks{This paragraph of the first footnote will contain the date on which you submitted your paper for review. It will also contain support information, including sponsor and financial support acknowledgment. For example, ``This work was supported in part by the U.S. Department of Commerce under Grant BS123456.'' }
\thanks{This work has been published in IEEE Transactions on Industrial Cyber-Physical Systems, DOI: 10.1109/TICPS.2024.3501278.}
\thanks{Z.~Qin is with the Automatic Control Laboratory, EPFL, Lausanne 1015, Switzerland. (email: zhaoming.qin@epfl.ch)}
\thanks{N.~Dong and Z.~Wang are with the Shanghai Artificial Intelligence Laboratory, Shanghai, 200232, China. (emails: \{dongnanqing,wangzhefan\}@pjlab.org.cn)}
\thanks{D. Liu is with the Department of Electrical Engineering, Tsinghua University, Beijing, 100084, China. (email: kfliudi@mail.tsinghua.edu.cn)}
\thanks{J.~Cao is with the Beijing National Research Center for Information Science and Technology, Tsinghua University, Beijing 100084, China. (email:  jcao@tsinghua.edu.cn)}
%\thanks{Corresponding authors: N.~Dong, D.~Liu,  and J.~Cao.}
% \thanks{This paragraph will include the Associate Editor who handled your paper.}
}

\maketitle


\begin{abstract}
% As a data-driven approach, multi-agent reinforcement learning (MARL) has made remarkable advances in solving the cooperative residential load scheduling problems. However, centralized training, the most commonly used paradigm for MARL algorithms, can raise privacy risks for the involved residents and preclude its large-scale deployment in the communication-restricted cloud-edge environments. In this work, we propose a novel multi-agent actor-critic framework where the global value function is approximated by the local critics in the edge layer and the feed-forward network in the cloud layer jointly. The residences in the edge layer only upload scalar individual value functions encoded by their local critics to the cloud, both preserving the privacy of residents and reducing the communication cost. The simulation experiments demonstrate that the proposed framework significantly outperforms the independent actor-critic framework, and can achieve comparable performance to the state-of-the-art actor-critic framework without privacy and communication constraints.
As a data-driven approach, multi-agent reinforcement learning (MARL) has made remarkable advances in solving cooperative residential load scheduling problems. However, centralized training, the most common paradigm for MARL, limits large-scale deployment in communication-constrained cloud-edge environments. As a remedy, distributed training shows unparalleled advantages in real-world applications but still faces challenge with system scalability, \emph{e.g.}, the high cost of communication overhead during coordinating individual agents, and needs to comply with data governance in terms of privacy. In this work, we propose a novel MARL solution to address these two practical issues. Our proposed approach is based on actor-critic methods, where the global critic is a learned function of individual critics computed solely based on local observations of households. This scheme preserves household privacy completely and significantly reduces communication cost. Simulation experiments demonstrate that the proposed framework  achieves comparable performance to the state-of-the-art actor-critic framework without data governance and communication constraints.
% the global value function is approximated by the local critics in the edge layer and the feed-forward network in the cloud layer jointly. The residences in the edge layer only upload scalar individual value functions encoded by their local critics to the cloud, both preserving the privacy of residents and reducing the communication cost. The simulation experiments demonstrate that the proposed framework significantly outperforms the independent actor-critic framework, and can achieve comparable performance to the state-of-the-art actor-critic framework without privacy and communication constraints.
\end{abstract}


\section{Introduction}
\label{}
As ultimate consumers in the electricity transmission chain, residential loads account for nearly 40\% of total electricity consumption in the developed countries (\emph{e.g.}, about 38.4\% in the U.S. in 2022 \cite{percent}).
The large flexibility of residential loads provides a great potential for energy regulation and scheduling, promoting the vigorous development of smart homes \cite{SHEM20}. By integrating multiple smart homes, the residential microgrid can aggregate the capacity of load scheduling and reduce the total energy costs.
So far, the load scheduling of the residential microgrid has gained increasing attention \cite{TIE_Wei2015,TIE_Wei17,MCTS21}.

In recent years, the breakthroughs in multi-agent reinforcement learning (MARL) have led to new solutions to the load scheduling problem \cite{Nature}. First, the residential microgrid with multiple households is naturally modeled as a multi-agent environment where each household is regarded as an agent. Second, without any prior knowledge of the residential microgrid, model-free reinforcement learning (RL) techniques can learn practical policies by interacting with the environment and then perform real-time execution based on the learned policies \cite{DRLSurvey}. Third, the emerging cloud-edge computing structure provides an ideal physical implementation for MARL \cite{MARL_cloud_edge}. %The centralized training is performed in the cloud layer, while decentralized execution is conducted in the edge layer.

Parallel to the design of MARL algorithms, another aspect to be considered is data governance, which is a collection of processes, policies, standards, and metrics that ensure the effective and efficient use of load scheduling data. Although the massive effort has been dedicated to developing the cooperative load scheduling schemes using MARL~\cite{Zhang19,Lee20,LiangYu,XuXu,Chung21,App_MAAC}, the privacy issues are tended to be ignored in these studies. Accessing household information during MARL training may breach user privacy. For example, residents' behaviors can be deduced from the arrival and departure times of household electric vehicles (EVs), and temperature preferences  can be inferred from the thermal comfort constraints \cite{TSGQin}. Since user data may contain sensitive information, strict data regulations have been established to ensure data governance~\cite{gdpr}.
Therefore, it is essential to develop a practical MARL framework to address the cooperative load scheduling problem in the cloud-edge environments while complying with data governance.

\subsection{Literature Review}
\label{Review}
\subsubsection{Multi-Agent Reinforcement Learning}
\label{sec:review:marl}
Several MARL frameworks have been developed to date \cite{IQL,COMA}. A simple framework is to integrate all agents as a single agent with joint state space and action space, where a single-agent RL algorithm is applied \cite{Du21}. For instance, a \textit{centralized actor-critic} (CAC) framework using prioritized deep deterministic policy gradient (DDPG) is employed to manage all devices of a residential multi-energy system \cite{CentralAppl}. Although this fully centralized framework theoretically allows cooperative behaviors across individual agents, it fails on simple cooperative MARL problems due to \textit{lazy} agents in practice \cite{VDN}. Furthermore,  the joint action space expands exponentially as the number of agents increases, leading to poor scalability \cite{QMIX}. Additionally, the centralized paradigm requires collecting local observations from all agents during the online execution phase, which imposes high demands on real-time communication. A remedy is \textit{distributed actor-critic} framework \cite{Distributed_AC}, in which each agent can access the observations of its neighbors via distributed communications networks.

In this work, we pursue fully decentralized policies depending only on local observations of agents. A straightforward framework to learn decentralized policies is to train the agents independently \cite{IQL}. For example, an \textit{independent actor-critic} (IAC) framework using the proximal policy optimization (PPO) algorithm is adopted to optimize a multi-household energy management scheme \cite{Zhang19}, while the independent Q-learning is applied to the demand response programs for different components in residential buildings \cite{IndependentAppl}. From the perspective of a single agent in such a framework, the behaviors and policies of other agents are not observable \cite{COMA}. Consequently, even if an agent's own policy remains unchanged, the global reward function it receives varies with the policies of other agents. Thus, each agent interacts with a non-stationary environment, making the learning process highly unstable.

To address the non-stationary environment during the learning process of decentralized policies, the framework \textit{decentralized actors with centralized critic} (DACC) is widely adopted by previous MARL approaches \cite{COMA,MADDPG,MAAC,DOP}. Its centralized critic has access to all agents' information during learning, ensuring accurate evaluation on the global rewards. While, the decentralized actors, \emph{i.e.}, the decentralized policies, are executed using their corresponding agents' information. Thus, DACC mitigates the challenge of non-stationary environments during learning. %For example, Chung \emph{et al.}~\cite{Chung21} used it to learn the cooperative load scheduling of multiple households. 
Nevertheless, these advantages of DACC come at the expense of agent privacy, since all agents' information must be shared with the centralized critic during learning.

Although few MARL algorithms take privacy issues into account \cite{Liu24}, some have the potential to preserve agents' private information during the training phase. For instance, in the multi-actor-attention-critic (MAAC) algorithm \cite{MAAC}, the original observations and actions are first encoded into embeddings by local functions. This approach ensures that the central attention accesses only the encoded data. Ye \emph{et al.}~\cite{App_MAAC} leveraged this characteristic to protect consumer privacy in local electricity markets. However, MAAC does not perfectly guarantee privacy, as the high-dimensional embeddings of agents remain vulnerable to privacy attacks. These embeddings, despite being encoded representations, could be exploited by adversaries via advanced machine learning techniques, such as deep learning-based inversion attacks or statistical inference methods. %By uncovering hidden patterns and correlations within these embeddings, attackers might deduce partially the original observations or actions, thereby compromising the agents' privacy. 
Making matters even worse, in the case that attackers gain access to the detailed parameters or structures of the local embedding functions,  the original observations and actions could be (partially) reconstructed,  posing a serious threat to agents' sensitive information including position trajectories and behavioral preferences.
Finally, it is impractical to deploy MAAC on a large scale in communication-restricted cloud-edge environments. The transmission of high-dimensional information between agents and the cloud significantly increases the communication burden in large-scale systems. Furthermore, the inherent self-attention mechanism incurs noticeable computational costs as the number of agents increases. 
% the access to full state,

\subsubsection{Data Governance}
\label{sec:review:dg}
Data governance is a data management concept focused on ensuring the availability, usability, integrity, and security of the data~\cite{khatri2010designing}. In the era of information technology, data privacy has emerged as a prominent topic of ethical and legal discussion~\cite{mehmood2016protection}. This work specifically addresses data privacy issues within the scope of data governance. Due to the risk of information leakage, data regulations~\cite{gdpr} prohibit the data holder from transferring user data out of local devices in any form~\cite{dong2022federated}. Note that this constraint differs from the privacy-persevering techniques widely adopted in the literature, such as differential privacy~\cite{abadi2016deep}. While integrating privacy-persevering techniques with MARL can safeguard user privacy in data transmission, it does not necessarily fulfill the requirements of data governance. 
% Thus, the discussion on privacy-persevering techniques is beyond the scope of this work.
\subsubsection{Edge artificial intelligence}
Recent research has introduced various technologies to enhance the efficiency and security of MARL in edge computing environment. Chen et al. explored edge multi-task transfer learning, providing valuable insights into data-driven task allocation methods crucial for optimizing multi-agent environments \cite{chen2019edge}.  Xiong et al. proposed a RL-based framework that intelligently allocates resources across edge devices, with the goal of improving key performance indicators such as latency  and energy efficiency \cite{xiong2020resource}. %By utilizing the adaptive learning capabilities of RL, this approach effectively addresses the complexity and dynamic nature of multi-agent environments, resulting in improved resource management and system scalability. 
These studies underscore the importance of combining advanced RL techniques with robust data governance frameworks to ensure privacy and efficiency of residential microgrids.

\subsection{Contributions}
% TODO On-policy & off-policy
In this work, we intend to minimize the total operation costs of a residential microgrid within a communication-restricted cloud-edge environment, while effectively preserving local household information. We formulate this cooperative load scheduling problem as a finite-horizon decentralized partial observable Markov decision process (Dec-POMDP)%\cite{DecPOMDP}
. To facilitate collaborative control of distributed demand-side resources and ensure user privacy, we introduce \textit{decentralized actors with distributed critics} (DADC), a novel MARL framework designed with data governance in mind. In the proposed framework, each household operates an individual actor and critic within the edge layer, relying solely on its local information. The local critic networks compute scalar value functions for each household, which are then transmitted to the cloud layer. By restricting communication to scalar values rather than exchanging raw data or model parameters, the framework ensures robust data governance. At the cloud layer, a global value function is estimated using a feed-forward network, which takes as input  the concatenation of all individual value functions. This hierarchical learning structure preserves privacy while enabling global optimization. 
The learning process for both the distributed critics and the cloud-level network is achieved by backpropagating the gradients derived from global temporal-difference (TD) updates, which are computed in the cloud layer based on the global reward signal. %Put differently, each individual critic is learned implicitly rather than from any reward specific to it. In this way, DADC efficiently addresses the credit assignment problem. 
This method effectively balances privacy, computational efficiency, and system performance, making it well-suited for decentralized optimization and deployment in resource-constrained edge environments.
The contributions of this paper are summarized as follows.\\
 1. We propose DADC, a novel MARL framework to address the cooperative load scheduling task of a residential microgrid while minimizing user privacy leakage.
  Unlike existing MARL frameworks used in most load scheduling schemes \cite{Zhang19,Lee20,LiangYu,XuXu,Chung21,Du21,CentralAppl, App_MAAC}, DADC ensures that each household only shares an encoded scalar value with the cloud layer during each time step of training phase, efficiently preserving private data. \\
2. In DADC, the global value function is computed using only the scalar individual values, and the cloud-level network enables linear computational complexity with respect to the number of households. These features facilitate the scalable deployment of DADC in the cloud-edge environments. \\
3. We empirically evaluate DADC using real-world load data, providing practical insights into the problem formulated in this work. Our results demonstrate that DADC significantly outperform IAC, a seminal baseline under data governance. Furthermore, DADC can achieve comparable performance to DACC, a general MARL framework without privacy-preserving mechanisms, highlighting its superiority in maintaining privacy while achieving efficient load scheduling. %And we investigate the data efficiency and learning efficiency of on-policy and off-policy training for DADC.

\section{Problem Formulation}

\subsection{Cloud-Edge Environment}
\begin{figure}[ht]
  \centering{\includegraphics[width=1.0\columnwidth]{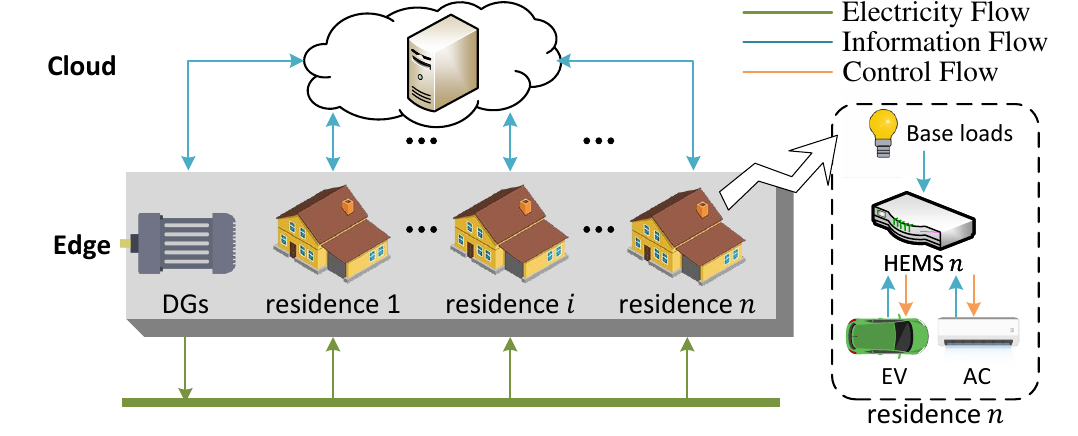}}
  \caption{The cloud-edge environment for residential load scheduling. The DGs supply electricity to households. The HEMSs take as input public information from the DGs and private observations from local households, and generate control signals for local flexible appliances.}
  \label{fig:system}
\end{figure}

Consider a cloud-edge environment for residential load scheduling. As illustrated in Fig. \ref{fig:system}, we assume an isolated microgrid at the edge layer, consisting of a set $\mathcal{D}=\{1,\ldots,n\}$ of $n$ households and distributed generators (DGs). The DGs and households communicate bidirectionally with cloud layer to maintain power balance and coordinate the operation of all flexible loads.
Without loss of generality, 
each household is assumed to be equipped with base loads, one EV and one air conditioner (AC).
Moreover, each household has one home energy management system (HEMS) that schedules controllable appliances including AC and EV.
To ensure data governance, each HEMS can only access public information from the DGs along with its own local information.
We consider this load scheduling problem over a horizon $T$ time steps, with each step of duration $\Delta t$, \emph{i.e.}, $t\in\mathcal{T}=\{1,\ldots,T\}$.

\subsection{System Model}

The ACs can be dynamically adjusted to maintain thermal comfort of the occupants in the corresponding households.
The indoor temperature dynamics for household $i$ are described as follows \cite{LiangYu},
\begin{equation}\label{Tdynamics}T_{i,t+1}^\mathrm{in}=\mathbf{F}_i^\mathrm{AC}\left(T_{i,t}^\mathrm{in},T^\mathrm{out}_{i,t}, P^\mathrm{AC}_{i,t},\varrho_{i,t}\right),   
\end{equation}
where $\mathbf{F}_i^\mathrm{AC}(\cdot)$ denotes the transition function of indoor temperature with respect to four variables, \emph{i.e.}, current indoor temperature $T_{i,t}^\mathrm{in}$, outdoor temperature $T^\mathrm{out}_{i,t}$, AC power $P^\mathrm{AC}_{i,t}$, and disturbance $\varrho_{i,t}$. %We note that the representation \eqref{Tdynamics} is general, incorporating common first-order thermodynamic model.
Accurately modeling thermal dynamics is typically intractable. 
Therefore, we assume that the explicit form of the function $\mathbf{F}_i^\mathrm{AC}(\cdot)$ is unknown.
The working power of ACs can be continuously adjusted within a range, 
\begin{equation}\label{PACcons}
  0\leq P^\mathrm{AC}_{i,t}\leq \overline{P}^\mathrm{AC}_{i},
\end{equation}
where $\overline{P}^\mathrm{AC}_{i}$ denotes the maximum working power of the AC in household $i$.
To ensure the thermal comfort of occupants, the following indoor temperature constraint should be satisfied, 
\begin{equation}\label{Tcons}
  \underline{T}_{i}^\mathrm{in}\leq T_{i,t}^\mathrm{in}\leq \overline{T}_{i}^\mathrm{in},  
\end{equation}
where $\underline{T}_{i}^\mathrm{in}$, $\overline{T}_i^\mathrm{in}$ denote the lower and upper limits of comfortable temperature in household $i$, respectively. 

%We assume that EVs can be charged or discharged during parking time. 
The dynamics of EV battery energy are as follows,
%$I_{i,t}$ is an indicator which is 1 when AC $i$ is used by occupant at time slot $t$ and 0 otherwise. 
%F_i^{EV}\left(E^i_{t},P^{EV}_{i,t}\right)
\begin{equation}\label{Edynamics}
  E^\mathrm{EV}_{i,t+1}=
    \begin{cases}
      E_i^\mathrm{init},&\mbox{if} \ t{+}1{=}t_i^\mathrm{a},\\
      E^\mathrm{EV}_{i,t}{+}\eta_i^\mathrm{c}P^\mathrm{EV}_{i,t}\Delta t,&\mbox{if} \ t_i^\mathrm{a}{\leq}t{<}t_i^\mathrm{d} \mbox{ and } P^\mathrm{EV}_{i,t}{\geq}0,\\
      E^\mathrm{EV}_{i,t}{+}P^\mathrm{EV}_{i,t}\Delta t/\eta_i^\mathrm{d},&\mbox{if} \ t_i^\mathrm{a}{\leq}t{<}t_i^\mathrm{d} \mbox{ and } P^\mathrm{EV}_{i,t}{<}0,\\
      0,&\mbox{otherwise}.\\
    \end{cases}
\end{equation}
In (\ref{Edynamics}), the variables $E^\mathrm{EV}_{i,t}$ and $P^\mathrm{EV}_{i,t}$ stand for the battery energy and the charging/discharging power of the EV in household $i$ at time step $t$, respectively. The parameters $E_i^\mathrm{init}$, $\eta^\mathrm{c}_i$, $\eta^\mathrm{d}_i$, $t_i^\mathrm{a}$ and $t_i^\mathrm{d}$ are the initial battery energy, the charging and discharging efficiency coefficients, the arrival time and departure time of the EV in household $i$, respectively. 
%$F_i^{EV}\left(\cdot\right)$ denotes the transition function of battery energy with respect to battery energy and charging/discharging power during parking time of EV at SH $i$.  
The target EV battery energy should be satisfied at the departure time of the EV, 
\begin{equation}\label{EVtask}
  E^\mathrm{EV}_{i,t^i_d}  \geq E_{i}^\mathrm{targ},
\end{equation}
where $E_{i}^\mathrm{targ}$ denotes the target battery energy of the EV in household $i$. 
Moreover, the charging/discharging power and the battery energy of the EV must be maintained within a range,
\begin{equation}\label{EVcons}
  -\overline{P}^\mathrm{EV}_{i}\leq P^\mathrm{EV}_{i,t}\leq \overline{P}^\mathrm{EV}_{i},\ \underline{E}^\mathrm{EV}_{i}\leq E^\mathrm{EV}_{i,t}\leq \overline{E}^\mathrm{EV}_{i},
\end{equation}
where $\overline{P}^\mathrm{EV}_{i}$, $\underline{E}^\mathrm{EV}_{i}$ and $\overline{E}^\mathrm{EV}_{i}$ represent the maximum charging/discharging power, the minimum and maximum battery energy of the EV in household $i$.

We assume that DGs have sufficient generation capacity to maintain the power balance of the whole microgrid. Moreover,
at each time step $t$, DGs are automatically adjusted to meet residential electricity needs,
\begin{equation}\label{balance}
  P_t^\mathrm{DG}=\sum_{i\in\mathcal{D}}\left( P^\mathrm{BL}_{i,t} + P^\mathrm{AC}_{i,t} + P^\mathrm{EV}_{i,t}\right),  
\end{equation}
where $P^\mathrm{BL}_{i,t}$ denotes the power of base loads in household $i$ at time step $t$.

\subsection{Objective Function} 

The total operation cost of the microgrid can be divided into two parts, \emph{i.e.}, the generation cost of DGs and the adjustment cost of DGs. 
The former is determined by the output power of DGs. 
The latter depends on the fluctuation of the output power of DGs
because the frequent power adjustment would degrade the service life of DGs.
Thus, the total cost at time step $t$ can be presented as follows. 
\begin{equation}\label{cost}
  C_t=\mathbf{G}_1\left(P_t^\mathrm{DG}\right)+\mathbf{G}_2\left(P_t^\mathrm{DG}-P_{t-1}^\mathrm{DG}\right),  
\end{equation}
where $\mathbf{G}_1(\cdot)$ is the generation cost function of DGs with respect to current output power of DGs \cite{7323868,7105897}, 
and $\mathbf{G}_2(\cdot)$ is the adjustment cost function of DGs with respect to the difference between the output power of DGs at current and last time step. At each time step $i$, the DGs report the incurred cost to the cloud.
It is notable that the functions $\mathbf{G}_1(\cdot)$ and $\mathbf{G}_2(\cdot)$ can be non-linear, thus the individual cost functions specific to households are not available in general. % DGs needs to satisfy the power ramping constraint:
%\begin{equation}
%  P_{min}^\mathrm{DG}\leq P_t^\mathrm{DG}-P_{t-1}^\mathrm{DG}\leq P_{max}^\mathrm{DG},  
%\end{equation}
%where $P_{min}^\mathrm{DG}$ and $P_{max}^\mathrm{DG}$  are the minimum and maximum ramping power of DGs, respectively.

Based on the above-mentioned models and objective function, a stochastic optimization problem 
minimizing the long-term microgrid operation cost can be formulated as follows,
\begin{equation} \label{eq:problem}
  \begin{split}\min_{P_{i,t}^\mathrm{AC},P_{i,t}^\mathrm{EV},i\in\mathcal{D},t\in\mathcal{T}} \quad &\mathbb{E}\left[\sum_{t\in\mathcal{T}} C_t\right]
      \\
    \mathrm{s.t.} \quad  &\eqref{Tdynamics}-\eqref{cost} 
   \end{split}
\end{equation}

\subsection{Dec-POMDP Formulation}

%At each timestep sequentially chooses actions for agents given all previous actions. 
In this subsection, we formulate the cooperative load scheduling problem following Dec-POMDP. The agents in Dec-POMDP are specified as the HEMSs in the microgrid. Since model-free MARL does not rely on the prior knowledge of state transition probability distribution, we focus on three components, the global state and local observations, the actions, and the global reward function.

\subsubsection{Dec-POMDP}
Let $\mathcal{P}(\Omega)$ denote the set of all probability distribution over the space $\Omega$. A finite-horizon Dec-POMDP \cite{QMIX} can be mathematically described by a tuple 
$\left<\mathcal{D},\mathcal{S},\mathcal{A},\mathcal{O},\mathbf{P}_s,\mathbf{P}_o,\mathbf{R},T\right>$, 
where 
\begin{itemize}
  \item $\mathcal{D}$ denotes the set of HEMSs. 
  \item $\mathcal{S}$ denotes the space of global states.
  \item $\mathcal{A}\equiv \times_{i\in\mathcal{D}}\mathcal{A}_i$ denotes the set of joint actions. 
  \item $\mathcal{O}\equiv \times_{i\in\mathcal{D}}\mathcal{O}_i$ denotes the set of joint observations.
  \item $\mathbf{P}_s:\mathcal{S} \times \mathcal{A} \mapsto\mathcal{P}(\mathcal{S})$ denotes the state transition function.
  \item $\mathbf{P}_o:\mathcal{S} \times \mathcal{A}\mapsto \mathcal{P}(\mathcal{O})$ denotes the joint observation function.
  \item $\mathbf{R}:\mathcal{S}\times \mathcal{A} \times \mathcal{S}\mapsto \mathbb{R}$ denotes the global reward function.
  \item $T\in \mathbb{N}^+$ denotes the horizon.
\end{itemize}
%$\mathcal{P}\left(\mathcal{X}\right)$ denotes the
%set of probability distributions over the set $\mathcal{X}$. 

At every time step $t$, each HEMS $i$ takes an action $a_{i,t}$ from its individual action space $\mathcal{A}_i$,
forming a joint action $\boldsymbol{a}_t$
which leads to a transition to a new state $\boldsymbol{s}_{t+1}{\sim}  \mathbf{P}_s(\boldsymbol{s}_t,\boldsymbol{a}_t)$ 
and a global reward $r_{t} = \mathbf{R}(\boldsymbol{s}_t,\boldsymbol{a}_t,\boldsymbol{s}_{t+1})$. 
Moreover, the environment emits a joint observation $\mathbf{o}_{t+1}\sim \mathbf{P}_o(\boldsymbol{s}_{t+1},\boldsymbol{a}_t)$ 
where each HEMS $i$ only draws its own observation $o_{i,t+1}$. 

The goal of the finite-horizon Dec-POMDP is to learn a joint policy 
$\bm{\pi}:\mathcal{O}\times\mathcal{A}\mapsto[0,+\infty)$ mapping a joint observation to a probability distribution over actions in \textit{continuous} joint action space, which maximizes the expectation of accumulated global reward 
$\sum_{t=1}^{T} r_{t}$. %We assume that the joint policy is stochastic. The notation of joint policy would degenerate into $\bm{\pi}:\mathcal{O}\mapsto\mathcal{A}$ if the policy is deterministic.
%\subsubsection{Value Functions and Advantage Function}
To evaluate the performance of the joint policy $\bm{\pi}$, we define the joint state value function as
\begin{equation}
  V^{\bm{\pi}}(\boldsymbol{o}_t):=\mathbb{E}_{\boldsymbol{s}_{t+1:T},\mathbf{o}_{t+1:T},\boldsymbol{a}_{t:T}\sim \bm{\pi}} \Big[\sum_{t'=t}^{T} r_{t'}|\boldsymbol{o}_t\Big].
\end{equation}
Here, $\mathbb{E}$ denotes the expectation.
The subscript of $\mathbb{E}$ enumerates the variables being integrated over,
where the global states, joint observations and actions are sampled sequentially from the dynamics model 
$\mathbf{P}_s$, $\mathbf{P}_o$ and policy $\bm{\pi}$, respectively.
Similarly, the joint state-action value function is defined as
\begin{equation}
  Q^{\bm{\pi}}(\boldsymbol{o}_t,\boldsymbol{a}_t):=\mathbb{E}_{\boldsymbol{s}_{t+1:T},\mathbf{o}_{t+1:T},\boldsymbol{a}_{t+1:T}\sim \bm{\pi}} \Big[\sum_{t'=t}^{T}  r_{t'}|\boldsymbol{o}_t,\boldsymbol{a}_t\Big].
\end{equation}
The advantage function $A^{\bm{\pi}}(\mathbf{o}_t,\boldsymbol{a}_t):=Q^{\bm{\pi}}(\boldsymbol{o}_t,\boldsymbol{a}_t)-V^{\bm{\pi}}(\boldsymbol{o}_t)$ measures whether the action $\boldsymbol{a}_t$ is better than the default behavior of policy $\bm{\pi}$.

\subsubsection{Global state and local observations} In the considered residential load scheduling scenario, the global state $\boldsymbol{s}_t$ incorporates the information owned by all HEMSs and the information from DGs. Since MARL algorithms do not operates over the global state, we omit the mathematical expression of $\boldsymbol{s}_t$.
To enable the cooperative scheduling of all households, the power of DGs is viewed as common information and provided to each HEMS. 
The observation of HEMS $i \in \mathcal{D}$ is 
\begin{equation}\label{ob}
  o_{i,t} = \left[t,P^\mathrm{DG}_{t},P^\mathrm{BL}_{i,t},P^\mathrm{PV}_{i,t},T^\mathrm{out}_{i,t},T_{i,t}^\mathrm{in},E_{i,t}^\mathrm{EV},E_{i}^\mathrm{targ},t_{i}^\mathrm{d}\right], 
\end{equation}
Here, the observation $o_{i,t}$ includes current time step $t$, which enables the policies to adapt to time-dependent behaviors, such as outdoor temperature and EV arrival/departure time. The last seven components in (\ref{ob}) are local information of household $i$ which should be preserved.
%The total power mismatch of all users can be calculated by  
%\begin{IEEEeqnarray}{rCl}
%  P^{mis}_{t} = \sum_{i\in\mathcal{N}} P^{B}_{i,t} + P^{AC}_{i,t} + P^{EV}_{i,t} - P^{PV}_{i,t},
%\end{IEEEeqnarray}

\subsubsection{Actions}
The problem \eqref{eq:problem} involves two categories of decision variables, namely the working power of ACs and the charging/discharging power of EVs. To facilitate the training of MARL, we unify the continuous action spaces of all decision variables to $[-1,1]$ by introducing control signals for ACs and EVs. Moreover, by designing the control signals, we rule out the possibility of generating policies that cause the loss of occupant comfort.
% The relationship between the control signals decided by the HEMS and the working powers executed by home appliances is presented as follows.

%The feasible  which are by different, see \eqref{PACcons} and \eqref{EVcons}. To guarantee the thermal comfort of the occupants and satisfy the indoor temperature constraint (\ref{Tcons}),
The mapping between the working power and control signal of AC of household $i$ is designed as follows,
\begin{equation}\label{eq:AC_scheme}
  P^\mathrm{AC}_{i,t} = 
  \begin{cases} 
    \overline{P}^\mathrm{AC}_{i},  &\mbox{if } T_{i,t}^\mathrm{in}\geq \overline{T}_i^\mathrm{in}  , \\ 
    0,  &\mbox{if }T_{i,t}^\mathrm{in}\leq \underline{T}_i^\mathrm{in} ,\\ 
    0.5\overline{P}^\mathrm{AC}_{i} (u_{i,t}^\mathrm{AC}+1), &\mbox{otherwise}.\\
  \end{cases}
\end{equation}
Provided that $u_{i,t}^\mathrm{AC}\in[-1,1]$,  the working power $P^\mathrm{AC}_{i,t}$ is forced to range between $[0,\overline{P}^\mathrm{AC}_{i}]$. Furthermore, the scheme \eqref{eq:AC_scheme} ensures the priority of occupant thermal comfort: 
AC is forced to run at the maximum power $\overline{P}^\mathrm{AC}_{i}$ when the room temperature exceeds the upper limit of comfortable temperature $\overline{T}_i^\mathrm{in}$, and turn off when the room temperature is below the lower limit of comfortable temperature $\underline{T}_i^\mathrm{in}$. The power of AC can be adjusted only when the temperature constraint (3) is satisfied.
%ACs would take corresponding mode to ensure the thermal condition: run at the maximum power $\overline{P}^\mathrm{AC}_{i}$ if room is above desirable maximum temperature $\overline{T}_i^\mathrm{in}$; turn-off if current indoor temperature is below the desirable minimum temperature.

Considering that the EV charging task (\ref{EVtask}) should be completed before the departure time, 
the following inequality must be checked for each time step $t_i^\mathrm{a}\leq t<t_i^\mathrm{d}$,
\begin{equation}\label{cheak}
  E_{i,t}^\mathrm{EV} + \eta_i^c  \overline{P}^\mathrm{EV}_{i}\left(t_i^\mathrm{d}-t\right)\Delta t \geq E_i^\mathrm{targ},
\end{equation}
where the left part indicates the EV battery energy at departure time if the EV is charged at maximum charging power during remaining charging time.
Once inequality (\ref{cheak}) is not satisfied, the corresponding EV is forced to be charged at maximum power.
Therefore, the following EV charging scheme during the charging time is formulated,
\begin{equation}\label{EVaction}
  P^\mathrm{EV}_{i,t} = 
  \begin{cases} 
    \overline{P}^\mathrm{EV}_{i},  &\mbox{if (\ref{cheak}) not satisfied}    , \\ 
    \max\{0,\overline{P}^\mathrm{EV}_{i}u_{i,t}^\mathrm{EV}\},  &\mbox{else if }  E^\mathrm{EV}_{i,t}\leq \underline{E}^\mathrm{EV}_{i}  , \\ 
    \min\{0,\overline{P}^\mathrm{EV}_{i}u_{i,t}^\mathrm{EV}\},  &\mbox{else if } E^\mathrm{EV}_{i,t}\geq \overline{E}^\mathrm{EV}_{i},\\ 
    \overline{P}^\mathrm{EV}_{i}u_{i,t}^{EV}, &\mbox{otherwise}.\\
  \end{cases}
\end{equation}
The $2^\textrm{nd}$ and $3^\textrm{rd}$ conditions in (\ref{EVaction}) guarantees that the battery energy of EV in household $i$ satisfies the constraint (\ref{EVcons}). 

Finally, the individual action of HEMS $i$ at time step $t$ is presented as $a_{i,t}=\left[u_{i,t}^\mathrm{AC},u_{i,t}^\mathrm{EV}\right]\in\left[-1,1\right]^2, i\in\mathcal{D}$.
The joint action formed by individual actions of all HEMSs at time step $t$ is
$\boldsymbol{a}_{t}=\left[a_{1,t},\ldots,a_{n,t}\right]\in\left[-1,1\right]^{2n}.$

\subsubsection{Reward}
The load scheduling problem intends to minimize the total operation cost, while the goal of Dec-POMDP is to maximize the accumulated reward. Therefore, we define the immediate global reward taking joint action $\boldsymbol{a}_t$ in state $\boldsymbol{s}_t$ as the negative cost of DGs, i.e., $r_{t} = -C_t$.

\section{MARL Framework in Cloud-Edge Environment under Data Governance}

In this section, we introduce DADC, a novel actor-critic framework designed to enable HEMSs to strictly preserve local observations while facilitating efficient collective training. 
We then elaborate the distributed training process for DADC, utilizing the PPO algorithm.

\subsection{Architecture}
\begin{figure}[t]
  \centering{\includegraphics[width=.9\columnwidth]{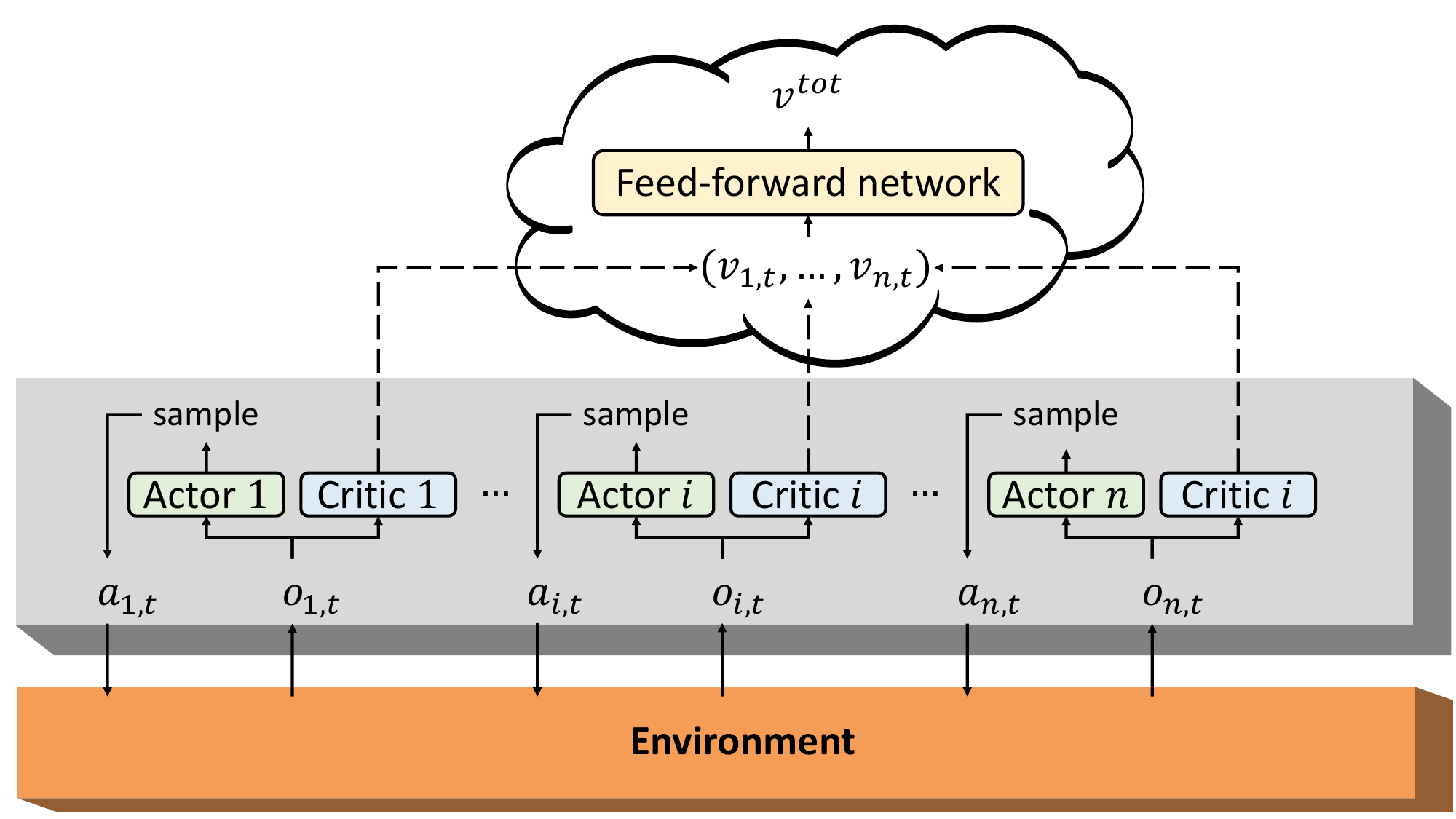}}
  \caption{The framework of DADC in the cloud-edge environment. At the edge layer, the actor network and critic network of each HEMS yield individual policy and a scalar value $v_{i,t}$ only using its local observation $o_{i,t}$, respectively. The individual action $a_{i,t}$ is then sampled according to the generated policy, and the scalar value  $v_{i,t}$ is communicated to cloud. At the cloud layer, a learnable feed-forward network maps the concatenation of $n$ scalar values to the global value estimation $v^{\mathrm{tot}}$.}
  %Each individual actor determines its policy only conditioning on its local observation. The global value function is computed
  \label{fig:network}
\end{figure}

DADC employs a structure consisting of decentralized actors and distributed critics. The critics are capable of estimating both the state value function and the action-state value function. For demonstration purposes, Fig.~\ref{fig:network} illustrates the structure of DADC, with critics approximating the state value function. The details of this structure are elaborated  below.   
\subsubsection{Decentralized Actors}
Each HEMS learns a stochastic policy $\pi_i:\mathcal{O}_i\times\mathcal{A}_i\mapsto [0,+\infty)$, parameterized by $\theta_i$, which maps its local observation to a probability distribution over its continuous action space.
Note that the policy $\pi_{i}$ is  conditioned only on the local observation $o_i$.
The joint policy $\bm{\pi}$ is then constructed from the decentralized policies $\{\pi_i\}_{i=1}^n$:
%The individual actor  denotes the stochastic policy of HEMS $i$.
%Let $\pi=\{\pi^1,\ldots,\pi^n\}$ denote the joint policy, which means
\begin{equation}
  \bm{\pi}\left(\boldsymbol{a}_t|\mathbf{o}_t\right):= \prod_{i=1}^n \pi_i(a_{i,t}|o_{i,t};\theta_i),
\end{equation}
where $\boldsymbol{a}_t=(a_{i,t},...,a_{i,t})$ and $\boldsymbol{o}_t=(o_{i,t},...,o_{i,t})$.

\subsubsection{Distributed Critics} DACC adopted by existing cooperative multi-agent actor-critic algorithms has a centralized critic to approximate the global value function of the joint policy, which requires the global state including the local observations of all agents, although the decentralized execution is allowed after training. To ensure data governance, the proposed DADC decomposes the approximation of the global value function into two steps,
\begin{subequations}\label{global_value_func}
\begin{align}
&v_{i,t} = V_{i}(o_{i,t};\phi_i),i=1,\ldots,n,\label{eq:v_i}\\
  &V^{\bm{\pi}}(\mathbf{o}_t)\approx V^{\mathrm{tot}}\left(v_{1,t},...,v_{n,t};\varphi\right)\label{eq:v_tot}.
\end{align}
\end{subequations}
In \eqref{eq:v_i}, individual critic $V_i(\cdot;\phi_i):\mathcal{O}_i\rightarrow \mathbb{R}$, parameterized by $\phi_i$, maps local observation $o_{i,t}$ to a scalar value $v_{i,t}$ at the edge layer. Subsequently, each HEMS transmits $v_{i,t}$ to the cloud. In \eqref{eq:v_tot},   a feed-forward network $V^{\mathrm{tot}}(\cdot;\varphi):\mathbb{R}^n\rightarrow \mathbb{R}$,  parameterized by $\varphi$, maps the concatenation of received $n$ scalar values to the global value estimation at the cloud layer.
Put differently, the approximation of the global value function at the cloud layer only requires the collection of the individual value functions, which are scalar values encoded by individual critics of agents at the edge layer. This design brings three key advantages. 
\begin{itemize}
    \item The local observations of agents are preserved strictly since it is intractable to analyze or deduce the original information through a scalar.
    \item The communication burden between the cloud layer and the edge layer is significantly reduced. For instance, when comparing DADC to MAAC, The local embedding functions in MAAC send transmit vectors of dimension  $d$ to the central attention mechanism. Thus, the total communication complexity in a cloud-edge environment using MAAC is $O(nd)$ whereas with DADC it is reduced to is $O(n)$.
    \item The computational burden at the cloud layer is also reduced by the use of a feed-forward network. The computational complexity is $O(n)$ if the hidden layers of the feed-forward network have fixed number of units, whereas MAAC's complexity is $O(n^2d)$ due to the self-attention network. 
\end{itemize}
Therefore, the proposed DADC framework facilitates large-scale deployment in the cloud-edge environment.

\subsubsection{Inner Structure}
\begin{figure}[ht]
\centering{\includegraphics[width=.9\columnwidth]{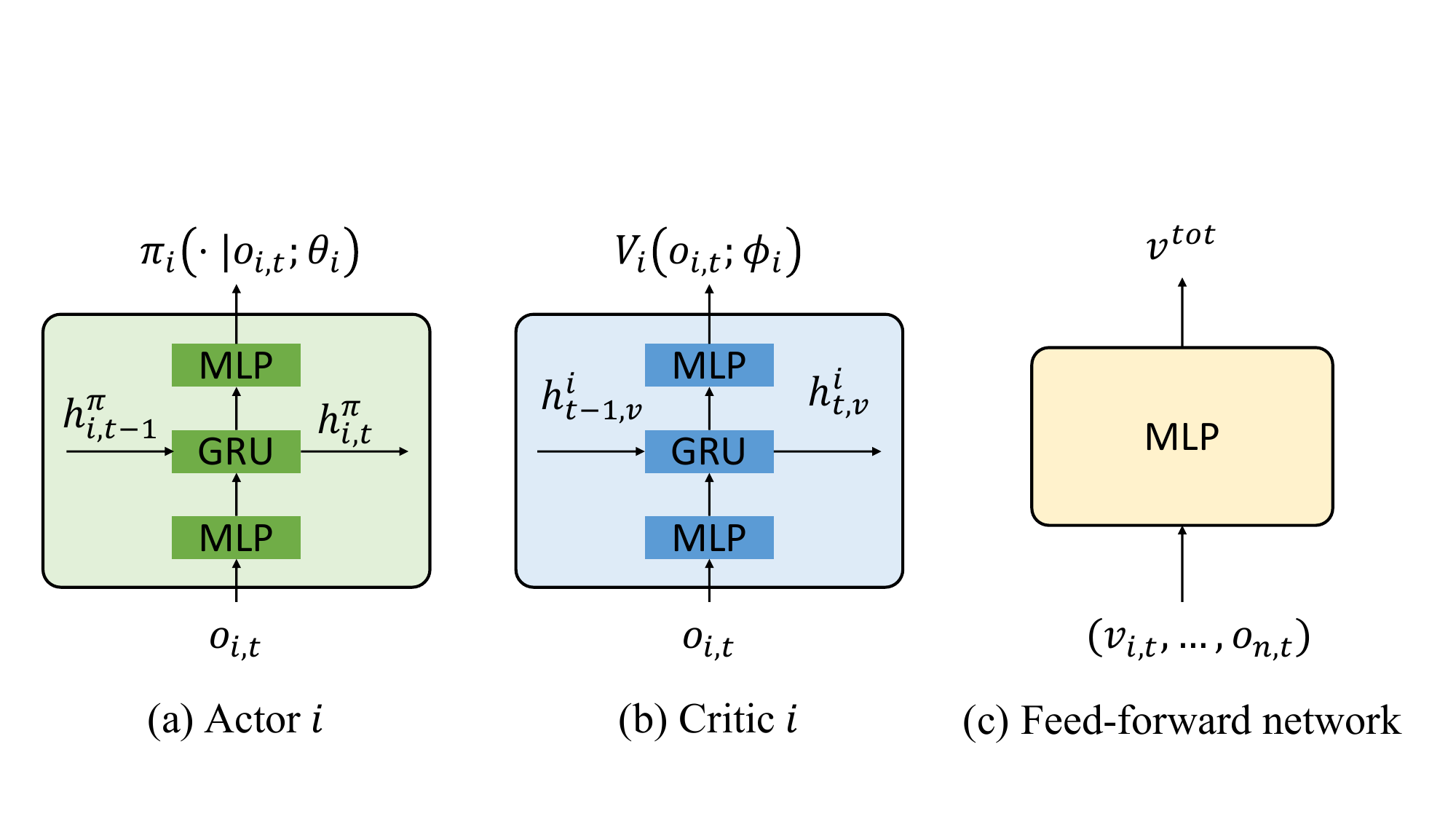}}
  \caption{(a) Individual actor network. This network takes as input the local observation $o_{i,t}$ and the hidden state  $h^\pi_{i,t-1}$, and generates the probability distribution over the individual action space.
  (b) Individual critic network. This network takes as input the local observation $o_{i,t}$ and the hidden state $h^v_{i,t-1}$  as input, and yields the individual value estimation. (c) Feed-forward network. This network take as input the concatenation of $n$ scalars, and outputs the global value estimation.}
  \label{fig:individual}
\end{figure}

The individual actors and critics and the feed-forward network are illuminated in Fig.~\ref{fig:individual}. Each agent's policy is represented by a combination of a gate recurrent unit (GRU) and multi-layer perceptrons (MLPs). The GRU module, a gating mechanism in recurrent neural networks, uses the hidden state $h^\pi_{i,t-1}$ to retain information from previous time steps, thereby enabling the agent to mitigate the challenges posed by partial observability.
Therefore, the integration of a GRU module and MLPs endows the individual actor network with the potential to generate effective policies based on limit observations. 
Similar to the structure of the individual actor network, the individual critic network also incorporates a GRU module and two MLPs. The feed-forward network at the cloud layer is entirely composed of MLPs.

\subsection{Distributed Training}
DADC can be trained using various RL algorithms in a distributed manner.
For demonstration purposes, we employ the PPO algorithm \cite{PPO} as an example and adapt the single-agent PPO to the multi-agent settings within a cloud-edge environment.

We first recall the single-agent PPO with an actor $\pi$ and a critic $V$. Given a policy $\pi$ parameterized by $\tilde{\theta}$, a batch of samples can be obtained, and the estimate of the advantage function $\hat{A}_t$ can be computed by the general advantage estimation (GAE) method \cite{GAE}
\begin{equation*}
  \hat{A}_t := \sum_{t'=t}^{T}\lambda^{t'-t}(-V(\mathbf{o}_{t'};\tilde{\theta}_v)+r_{t'}+V(\mathbf{o}_{t'+1};\tilde{\theta}_v)),    
\end{equation*}
where parameter $\lambda$ is used to control the trade-off between variance and bias of the estimate, and $\tilde{\theta}_v$ is the parameter of the critic $V$. Then, the actor network is updated by minimizing the loss
\begin{equation*}
\mathcal{L}^a(\theta):=\hat{\mathbb{E}}_t[\min (w_t(\theta)\hat{A}_t,\text{clip} (w_t(\theta),1-\epsilon,1+\epsilon)\hat{A}_t)].
\end{equation*}
Here, the expectation $\hat{\mathbb{E}}_t[\cdot]$ denotes the empirical estimation over a finite batch of samples. 
The probability ratio $w_t(\theta)$ is defined as $\frac{{\pi}(\boldsymbol{a}_t|\mathbf{o}_t;\theta)}{{\pi}(\boldsymbol{a}_t|\mathbf{o}_t;\tilde{\theta})}$. The hyperparameter $\epsilon$ limits the change in the probability ratio. 
The critic network $V$ is updated by minimizing 
$\mathcal{L}^c(\theta_v) :=\hat{\mathbb{E}}_t [(V(\mathbf{o}_t;\theta_v)-V(\mathbf{o}_t;\tilde{\theta}_v)-\hat{A}_t)^2].$

In the DADC framework, the global value function is approximated by (\ref{global_value_func}). Consequently, the global critic loss in the multi-agent settings is computed as 
\begin{equation}
      \begin{split}
      \mathcal{L}^c := \hat{\mathbb{E}}_t \left[(V^{\mathrm{tot}}(v_{1,t},...,v_{n,t};\varphi)-\tilde{v}^\mathrm{tot}-\hat{A}_t )^2 \right].
  \end{split} 
\end{equation}
where $\tilde{v}^\mathrm{tot}:=V^{\mathrm{tot}}(v_{1,t},...,v_{n,t};
  \tilde{\varphi})$.
By applying the the chain rule, the gradient of the feed-forward network at the cloud layer is given by $\triangle\varphi=\frac{\partial\mathcal{L}^c}{\partial \varphi}$,
while the gradient of the individual critic $i$ at the edge layer is $\triangle\phi_i =\frac{\partial \mathcal{L}^c}{\partial V_i}\frac{\partial V_i}{\partial \phi_i}$.
Note that the gradient term $\frac{\partial \mathcal{L}^c}{\partial V_i}$ must be communicated from the cloud to HEMS $i$.
In this manner, each individual critic $V_i(\cdot;\phi_i)$ is trained by backpropagating gradients from the global TD updates, which depends on the joint global reward. In other words, $V_i(\cdot;\phi_i)$ is learned implicitly rather than from any reward specific to HEMS $i$.

The individual actor loss function is defined as 
\begin{equation}
\mathcal{L}^a_i(\theta_i)=\hat{\mathbb{E}}_t\left[\min (w_{i,t}(\theta_i)\hat{A}_t, \mathrm{clip} \left(w_{i,t}(\theta_i), 1{-}\epsilon,1{+}\epsilon\right)\hat{A}_t)\right], 
\end{equation}
where $w_{i,t}(\theta_i){:=}\frac{\pi_i(a_{i,t}|o_{i,t};\theta_i)}{\pi_i(a_{i,t}|o_{i,t};{\tilde{\theta}_i})}$. Note that the individual loss $\mathcal{L}^a_i(\theta_i)$ is calculated locally by HEMS $i$ once the global value function is received. Then, the gradient of individual actor $i$ can be computed as 
$\triangle\theta_i =\frac{\partial \mathcal{L}^a_i}{\partial \theta_i}$.

\begin{algorithm}
  \caption{Distributed Training for DADC with PPO}\label{alg:training}
  \begin{algorithmic}[1]
      \State Initialize $\theta_i$ and $\phi_i$ for each HEMS; initialize $\varphi$ for feed-forward network.
      \For{$\mathrm{episode}=1$ to $\mathrm{episode}_\mathrm{max}$}
        
        \State\tikzmk{A}\% Interact with the environment 
        %\hspace*{-\fboxsep}\colorbox{lightgray}{\parbox{\linewidth}{ 
        \For{$t=1$ to $T$} 
          \ForAll{HEMSs $i$}
            \State $\tilde{\theta}_i\leftarrow\theta_i$, $\tilde{\phi}_i\leftarrow\phi_i$
          \State Sample action $a_{i,t} \sim \pi_i(\cdot|o_{i,t};\tilde{\theta}_i)$.
            \State Execute action $a_{i,t}$ and observe $o^i_{t+1}$. 
            %\State Receive global reward $r_t$ from Cloud.
            \State $\tilde{p}_{i,t}\leftarrow\pi_i(a_{i,t}|o_{i,t};\theta_i)$, $\tilde{v}_{i,t}\leftarrow V_i(o_{i,t};\phi_i)$.
            \State \textbf{Upload} $\tilde{v}_{i,t}$ to cloud. \Comment{Comm.}
          \EndFor
        \EndFor
        
        %}}
        \State \tikzmk{B} \boxit{gray} 
        \tikzmk{A} \% Estimate global advantage function
        %\hspace*{-\fboxsep}\colorbox{Goldenrod}{\parbox{\linewidth}{ 
        \State $\hat{A}_T\leftarrow 0,\tilde{v}^{\mathrm{tot}}_{T+1}\leftarrow 0,\tilde{\varphi}\leftarrow\varphi$ 
        \For{$t=T$ to $1$}
          \State $\tilde{v}^{\mathrm{tot}}_{t}\leftarrow V^{\mathrm{tot}}(\tilde{v}_{1,t},...,\tilde{v}_{n,t};\tilde{\varphi})$
          \State $\hat{A}_t\leftarrow \lambda\hat{A}_{t+1}+r_t+\gamma \tilde{v}^{\mathrm{tot}}_{t+1}-\tilde{v}^{\mathrm{tot}}_{t}$
          %\State $\hat{A}_t\leftarrow \hat{R}_t$
        \EndFor
        
        \State{Send $\{\hat{A}_t\}_{t=1}^T$ to each HEMS. \Comment{Comm.}} 
        %}}
        \State \tikzmk{B} \boxit{orange} 
        \% Update parameter 
            \State \tikzmk{A}\%  Edge layer
            \ForAll{HEMSs $i$}
                \State $\{v_{i,t}\}_{t=1}^T\leftarrow \{V_i(o_{i,t};\phi_i)\}_{t=1}^T$
                \State \textbf{Upload} $\{v_{i,t}\}_{t=1}^T$.\Comment{Comm.}
            \EndFor
            \State\tikzmk{B}\boxitt{gray} \tikzmk{A}
             $\mathcal{L}^c\leftarrow \sum_{t=1}^T(V^{\mathrm{tot}}(v_{1,t},...,v_{n,t};\varphi)-\tilde{v}^{\mathrm{tot}}_{t}-\hat{A}_t)^2$
            \State Update $\varphi$ with gradient $\partial \mathcal{L}^c/ \partial \varphi$.
            \State \textbf{Send} $\{\partial \mathcal{L}^c/ \partial v_{i,t}\}_{t=1}^T$ to HEMS $i$. \Comment{Comm.}
            \State \tikzmk{B} \boxitt{orange}\tikzmk{A}\% Edge layer
            \ForAll{HEMSs $i$}
                \State $\triangle \phi_i\leftarrow \sum_{t=1}^T\partial \mathcal{L}^c/ \partial v_{i,t}\cdot \partial v_{i,t}/ \partial \phi_i$
                \State Update $\phi_i$ with gradient $\triangle \phi_i$.
                \For{$t=1$ to $T$}
                    \State $w_{i,t}\leftarrow \pi_i(a_{i,t}|o_{i,t};\theta_i)/\tilde{p}_{i,t}$
                \EndFor
                \State $\mathcal{L}^a_i \leftarrow\sum\min (w_{i,t}\hat{A}_t,\text{clip} (w_{i,t},1-\epsilon,1+\epsilon)\hat{A}_t)$
                \State $\triangle \theta_i \leftarrow \sum_{t=1}^T\partial \mathcal{L}^a_i/ \partial w_{i,t}\cdot \partial w_{i,t}/ \partial \theta_i$
                \State Update $\theta_i$ with gradient $\triangle \theta_i$.
                \State \tikzmk{B}\boxitt{gray}
            \EndFor
            
      \EndFor
  \end{algorithmic}
\end{algorithm}

The distributed training process in the cloud-edge environment is detailed in Algo.~\ref{alg:training}.
Each training iteration is divided into three primary stages, 1) interaction with the environment, 2) estimation of the global advantage function, and 3) parameter updates. In the algorithm,  operations executed at the edge layer are shaded gray, while those at the cloud layer are shaded yellow for clarity.

In the first stage, shown in lines 4-10, agents operates in a fully decentralized manner. 
At each time step, each HEMS interacts with the environment by independently selecting and executing its own action. The only information uploaded to the cloud layer is the individual value estimation calculated by each HEMS.

The cloud layer exclusively performs the second stage. As shown in line 14, the global value function is estimated using only the scalar value estimates received from HEMSs, without requiring access to their local observations or actions. The estimation of the global advantage function at time step $t$ employs a backward-view TD method, utilizing the advantage function at time step $t+1$.

The third state involves a collaborative update process between the edge and cloud layer. HEMSs first calculate their individual value functions using the updated parameters and transmit these scalar values to the cloud layer. The cloud computes the gradients of the global critic loss with respect to both the feed-forward network parameters and individual value functions. These gradients are used to update the feed-forward network and are distributed back to the respective HEMSs. Finally, each HEMS update its local actor and critic networks using the received gradients and global advantage functions.

\section{Experiments}
In this section, we present the simulation experiments and report the empirical results. We begin by detailing the experimental setup of simulations. Next, we compare the proposed DADC with existing actor-critic frameworks using PPO, and analyze the effectiveness of the learned policies in the context of cooperative load scheduling. Finally, we access the scalability of DADC by evaluating its performance across varying numbers of households.

\subsection{Experiment Setup and Implementation}

\subsubsection{Environment}
The simulation environment is built upon OpenAI Gym \cite{openai}. 
The dynamics functions, cost functions and important household parameters are provided in the Appendix.
We consider the energy management problem over a one-day period, using a time step of 15 minutes,
resulting in a time horizon of $T=96$. 
Real-world power assumption data and temperature data are employed to model the power
of basic loads and outdoor temperature, sourced from Pecan Street Database~\cite{Database} and NOAA~\cite{NOAA}, respectively. We first consider a standard scenario consisting of 10 heterogeneous households in the following three subsections. In Sec. IV-E, we investigate the performance of the proposed DADC in large-scale scenarios.

\subsubsection{Network Architecture}
The individual critic networks share the same structure, consisting of three components, as shown in Fig. \ref{fig:network}, a fully-connected MLP with two layers of 64 units followed by $tanh$ nonlinearity, 
a GRU layer with 64 units, and a fully-connected MLP with one hidden layer of 128 units and one ouput layer of 1 units. 

To represent the stochastic policy, we use a Gaussian distribution $\mathcal{N}(\mu,\sigma^2)$ in this work. 
Therefore, the individual actor networks have one $tanh$ output for the mean $\mu$ and another $sigmoid$ output for the variance $\sigma^2$. 
The non-output layers of individual actor networks share the same structure with the individual critic networks.

The MLP of the feed-forward network in the cloud layer is consisted of one hidden layer of 64 units followed by $tanh$ nonlinearity and one output layer of 1 units.
\subsubsection{Baseline Frameworks}
Two baseline frameworks are considered: IAC and DACC. Notably, one of primary goals of this study is to ensure data governance. In contrast to the proposed DADC and IAC, the DACC framework raises non-trivial concerns regarding both privacy and communication costs. Thus, we  report the performance of DACC only in Sec.~\ref{sec:exp:main} and Sec.~\ref{sec:exp:implicit} for quantitative comparison.

Under the \textbf{IAC} framework, each HEMS comprises an independent actor and critic, following the same architecture as the individual actor and critic in DADC. The actor and critic of each HEMS are trained using the single-agent PPO algorithm, thereby eliminating the need for communication among agents. 

The \textbf{DACC} framework maintains decentralized actors for agents and a centralized critic. The decentralized actors share the same network as the individual actors in DADC, while the centralized critic adopts the network shown in Fig. 3 (b), taking the joint observation of all HEMSs as input rather than the local observation of a single HEMS. 

\subsubsection{Shared Hyperparameters}
We optimize the actor and critic networks using Adam with the learning rate of $1\times10^{-4}$ and $3\times10^{-4}$, respectively. 
The network parameters are updated every 120 environment steps with the batch size of 120.
We run 10 parallel environments to improve the training efficiency.
The case studies are conducted on a server with an 8-core AMD Ryzen 7 3700X processor and one single GeForce RTX 2080 GPU.

\subsection{Algorithm performance} \label{sec:exp:main}
First, we compare the proposed DADC framework with other actor-critic frameworks on the  cooperative load scheduling problem. For a fair comparison, each frameworks is trained with PPO for six times with different random seeds. In PPO, the GAE parameter is set to be 0.95, and the network parameters are updated 3 times per sample~\cite{PPO}.

We apply the following evaluation procedure during training:  for each trial, training is paused every 1000 episodes, and 10 independent episodes are run with each agent performing decentralized action selection. 
The cumulative reward for each episode is termed the \textit{episode reward}.

\begin{figure}[htbp]
  \centering{\includegraphics[width=.8\columnwidth]{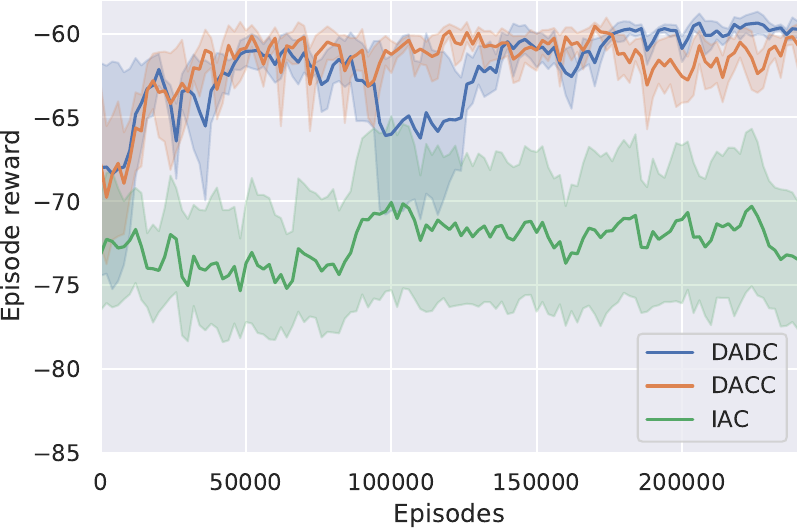}}
  \caption{Training curves of DADC and other frameworks. The solid curves corresponds to the mean and the shaded region to the minimum and maximum episode rewards over the all trials.}
  \label{fig:reward}
\end{figure}

The training curves are shown in Fig. \ref{fig:reward}. We observe that IAC fails to learn stable policies, resulting in poor performance, arguably due to the non-stationary environments encountered by its independent agents. In contrast, DACC leverage a global critic to facilitate more stable learning of coordinated behaviors across agents. DADC, on the other hand, achieves slightly better performance than DACC.  The policies of DADC escape the local minimum of DACC at the price of a sharp performance decline at about $1\times 10^5$ episodes. This implies that DADC has a better exploration capability.
Note that DADC preserves local information, unlike DACC. Thus, Fig.~\ref{fig:reward} demonstrates the superior performance of DADC over other actor-critic frameworks in this cooperative load scheduling task.

\subsection{Effect of Implicit Credit Assignment}
\label{sec:exp:implicit}

\begin{figure}[htbp]
  \centering{\includegraphics[width=0.85\columnwidth]{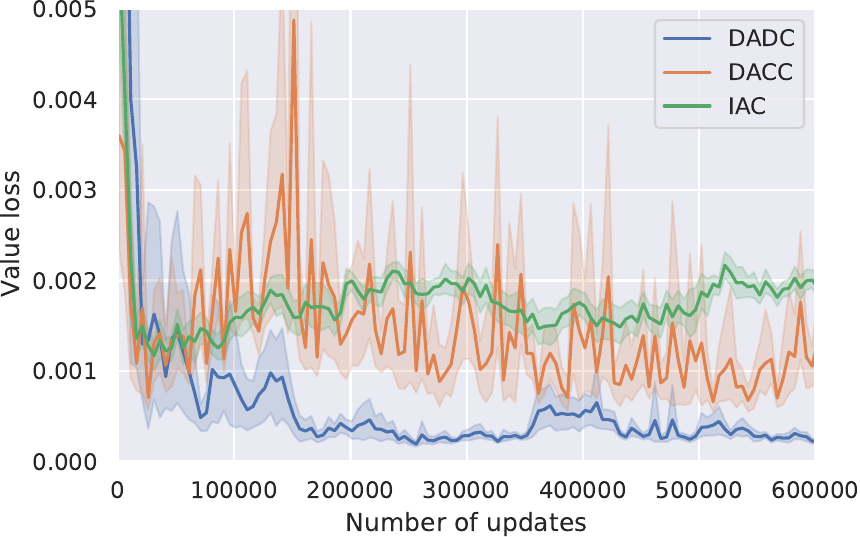}}
  \caption{The value loss for critic networks. DADC achieve the lowest estimation bias for global value function.}
  \label{fig:value_loss}
\end{figure}
As discussed in Section \ref{Review}, DACC can implicitly learn credit assignment across agents. To demonstrate this, 
we plot the value loss for critic networks in Fig.~\ref{fig:value_loss}. 
We observe that the independent critic networks in IAC exhibit the highest estimation bias, as the fully decentralized HEMSs in IAC cannot account for the dynamic behaviors of other HEMSs during training.

DACC, with its centralized critic, shows comparatively smaller value loss than IAC. However, the estimate of the global value function remains highly unstable during training. This instability arises because DACC's centralized critic processes observations from all HEMSs, making it slow to adapt to changes in the global reward when any single HEMS adjusts its policy.

In contrast, DADC enables  all HEMSs to cooperatively estimate the global value function through distributed critic networks, allowing each individual value function to be learned via end-to-end training.
In Fig.~\ref{fig:value_loss}, DADC achieves a much lower value loss than IAC and DACC, highlighting  the effectiveness of implicit credit assignment in DADC. This finding partially explains why DADC performs on par with DACC, despite the cloud receiving considerably less information from each household.

\subsection{Effect of Load Scheduling}

We next examine the control effects of DADC on the cooperative load scheduling task.
After training, we test the policies that achieved the best evaluation performance during the training phase.
The test results, shown in Table \ref{table:performance}, indicate that DADC reduces the average cost  by 11\% compared to IAC. 
Notably, the adjustment cost is reduced by more than 50\%.

\begin{table}[htbp]
  \scriptsize
  \centering
  \caption{Test performance for different actor-critic frameworks}    \label{table:performance}
  \begin{tabular}{cccc}
  %p{0.04\textwidth}>p{0.04\textwidth}>p{0.04\textwidth}>p{0.04\textwidth}>p{0.04\textwidth}>p{0.04\textwidth}
      \toprule
      \textbf{Metrics}&\textbf{DADC}&\textbf{DACC}&\textbf{IAC}\\
      \midrule
      Average Total Cost& $58.2\pm 0.9$& $65.4\pm 1.2$ &  $59.5\pm1.0$\\
       Average Generation Cost&$55.8\pm 1.0$&  $60.6\pm 1.5$ & $56.7\pm 1.1$\\
       Average Adjustment Cost&$\phantom{0}2.4\pm 0.3$&$\phantom{0}4.9\pm 0.8$ & $\phantom{0}2.8\pm 0.4$\\
      \bottomrule
  \end{tabular}
  \end{table}

\label{sec:exp:load}
 \begin{figure}[htbp]
  \centering{\includegraphics[width=0.8\columnwidth]{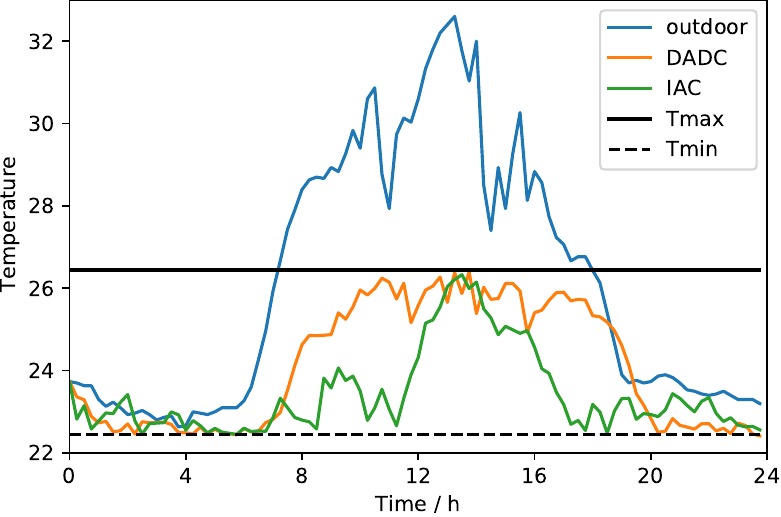}}
  \caption{Indoor temperature. The black solid and dashed lines denote the desirable maximum and minimum indoor temperature, respectively. The orange and green lines denote the indoor temperature curves during one day controlled by decentralized policies with DADC and IAC, respectively.}
  \label{fig:temp}
\end{figure}
To present the control effects for ACs, we plot the indoor temperature curves for a single AC over one day in Fig.~\ref{fig:temp}. 
Both frameworks control the indoor temperature within the specified constraints.
However, with DADC, the indoor temperature remains closer to the upper temperature constraint when outdoor temperatures are high, resulting in energy savings and  cost reduction compared to IAC.
Additionally, the indoor temperature curve with DADC is relatively smooth, 
indicating fewer adjustments to the AC than with IAC.

\begin{figure}[htbp]
  \centering{\includegraphics[width=0.8\columnwidth]{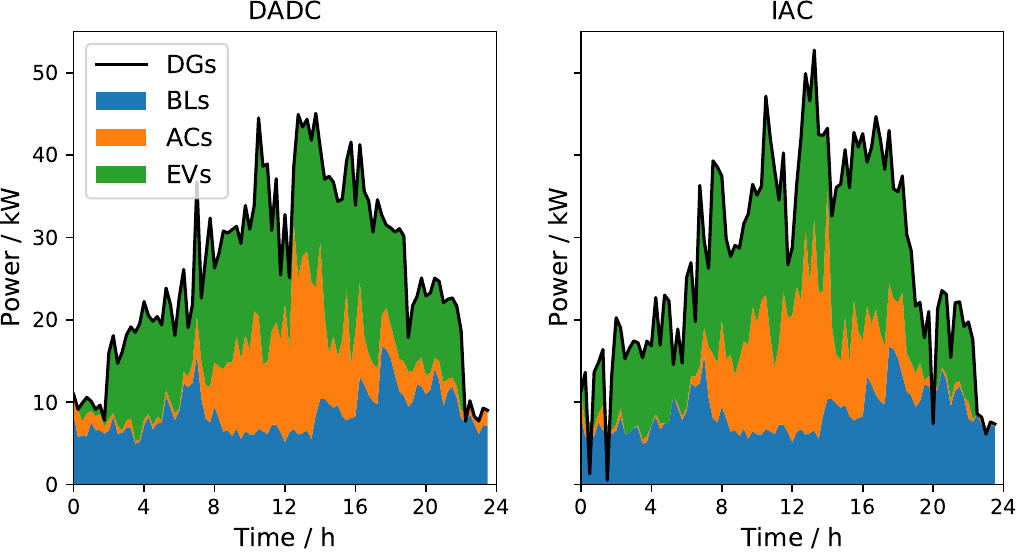}}
  \caption{Load scheduling. The blue, orange and green area denote the power of base load, the total power of ACs and the total charging power, respectively.}
  \label{fig:load}
\end{figure}
To demonstrate the overall load scheduling, 
Fig.~\ref{fig:load} displays the base load power, DGs power, total charging/discharging power of EVs, and total working power of ACs. Compared to IAC, the cooperative load scheduling achieved by DADC reveals two salient characteristics. First, the output power adjustments for DGs are relatively stable, resulting to  lower adjustment costs for DGs. Second, the peak power of DGs is lower than that with IAC. This indicates that DADC enables HEMSs to learn decentralized policies that allow households to cooperatively schedule load and reduce the global cost. In contrast, IAC fails to learn such  cooperative policies due to its fully independent actor-critic structure.

\subsection{Scalability Evaluation}
The above subsections presented and discussed simulation results for a scenario with 10 households. In this subsection, we empirically evaluate the effectiveness of the proposed DADC framework as the number of households increases. Table  \ref{table:scalability_DACC} and Table \ref{table:scalability_DADC} report two key metrics for DADC and DACC in scenarios with 10, 100 and 1000 households. The first metric, communication traffic between HEMSs and the cloud, grows linearly with respective to the number of households. The second metric, total processing time at the cloud layer, reflects the precessing burden on the cloud. 

Table  \ref{table:scalability_DACC} and Table \ref{table:scalability_DADC} show that DADC requires less than one-fifth of the communication overhead needed by DACC. Moreover, as the number of households increases, the cloud computational efficiency advantage of DADC over DACC becomes more pronounced. Thus, DADC shows significant scalability advantage over DACC in terms of both communication and cloud computational burdens.
\begin{table}[htbp]
  \centering
 \caption{Scalability evaluation for DACC}    \label{table:scalability_DACC}
  \begin{tabular}{cccc}
      \toprule
      \textbf{Number of households}&10 & 100 & 1000\\
      \midrule
      \textbf{Communication traffic (MB)}  & 16.9 & 169 & 1690 \\
       \textbf{Computation burden (hours)}  & 0.8 & 3.7 & 33\\
      \bottomrule
  \end{tabular}
  \end{table}
\begin{table}[htbp]
  \centering
 \caption{Scalability evaluation for DADC}    \label{table:scalability_DADC}
  \begin{tabular}{cccc}
      \toprule
      \textbf{Number of households}&10 & 100 & 1000\\
      \midrule
      \textbf{Communication traffic (MB)}  & 3.1 & 31 &310 \\
       \textbf{Computation burden (hours)} & 0.5 & 0.82 & 4.1\\
      \bottomrule
  \end{tabular}
  \end{table}
\section{Conclusion}
This paper proposes a novel multi-agent actor-critic framework, DADC, to address the cooperative load scheduling problem in a communication-restricted cloud-edge environment. 
A salient feature of DADC is its two-step approximation of the  global value function. First, each HEMS's individual critic network maps its local information into a scalar value, which is subsequently uploaded to the cloud. Second, the cloud estimate the global value function with a feed-forward network that takes these scalar values as inputs.

This framework brings three significant benefits. First, it enhance user privacy protection. Second, it significantly reduces communication traffic and computational burden on the cloud, thereby improving the training efficiency and scalability. Third, despite the cloud's access to limited information, the decentralized policies learned by HEMSs achieve performance comparable to that of DACC, arguably due to improved implicit credit assignment.

The current DADC framework assumes fixed load types and a fixed number of households during training, which restricts its applicability. Future work will focus on adapting DADC to general scenarios with varying load types  and dynamic household participation, enabling broader broader adoption of cooperative residential load scheduling.

\section*{Appendix} 

The transition functions and cost functions used for simulation are specified as follows.
\begin{equation}
  \mathbf{F}_i^{AC}(T,T^\mathrm{out},P,\varrho) = T+\alpha_i(T^\mathrm{out}-T)-\beta_iP+\varrho,
\end{equation}
where $\alpha_i$ and $\beta_i$ are the the coefficients associated with the
thermal characteristics of corresponding room and AC, and $\varrho$ follows the uniform distribution $\mathcal{U}[-0.1,0.1]$. The arrival time $t_i^\text{a}$ is a random variable with probability distribution $\mathcal{U}[\psi_i,\psi_i+\delta_1]$, where $\psi_i$ indicates  the parking habit of the occupant in household $i$, and $\delta_1$ is a shared parameter representing the variance of arrival time. Given the arrival time $t_i^\text{a}$, we assume the departure time $t_i^\text{d}\sim \mathcal{U}[t_i^\text{a}+\delta_2,t_i^\text{a}+\delta_3]$, implying that the dwell time of EV $i$ ranges between $[\delta_2,\delta_3]$. The parameters $\delta_1,\delta_2$ and $\delta_3$ are set to be $3,9$ and $12$, respectively.  Other parameters are randomly sampled according to the range in Table \ref{table:SH}.
\begin{table}[ht]
  % \scriptsize
  \centering
  \caption{Parameter ranges of households}    \label{table:SH}
  \begin{tabular}{cccccc}%p{0.04\textwidth}>p{0.04\textwidth}>p{0.04\textwidth}>p{0.04\textwidth}>p{0.04\textwidth}>p{0.04\textwidth}
      \toprule
      \textbf{Parameters} &$\underline{T}_i^\mathrm{in}$ & $\overline{T}_i^\mathrm{in}$ & $\overline{P}_{i}^\mathrm{AC}$  \\
      \textbf{Range} & $[22,24]$ & $[26,28]$ & $[3,4]$   \\
        \midrule 
      \textbf{Parameters} & $\alpha_i$ & $\beta_i$ & $\overline{P}_{i}^\mathrm{EV}$  \\   
      \textbf{Range} & $[0.19,0.21]$ & $[0.5,0.7]$ & $[6,10]$\\
        \midrule 
      \textbf{Parameters} & $\overline{E}_i^\mathrm{EV}$ & $\eta_i^\mathrm{c}$ & $\eta_i^\mathrm{d}$  \\   
      \textbf{Range} &  $[40,60]$ & $[0.90,0.95]$ & $[0.90,0.95]$ \\
      
      \bottomrule
  \end{tabular}
  \end{table}
The cost functions of DGs are specified as 
  \begin{equation}
  \begin{split}
      \mathbf{G}_1(P)&=\lambda_1^\mathrm{DG}P+\lambda_2^\mathrm{DG}P^2, \\
      \mathbf{G}_2\left(P_t,P_{t-1}\right)&= \lambda_3^\mathrm{DG}|P_t-P_{t-1}|.
  \end{split}
  \end{equation}
where $\lambda_1^\mathrm{DG},\lambda_2^\mathrm{DG},\lambda_3^\mathrm{DG}$ denote the cost coefficients of DGs, and are selected as 0.5, 0.0125 and 0.1, respectively.





\bibliographystyle{IEEEtran}
\bibliography{mybibfile}

\end{document}